\documentclass[showpacs,onecolumn]{revtex4}

\usepackage{ragged2e}
\usepackage{amsmath}
\usepackage{graphicx}
\usepackage{rotating}

\usepackage[usenames,dvipsnames]{color}

\definecolor{darkblue}{RGB}{0,0,196}

\begin{document}

\title{\Large \bf Excitation functions of related temperatures of
$\eta$ and $\eta ^0$ emission sources from squared momentum transfer spectra
in high-energy collisions\vspace{0.5cm}}

\author{Qi Wang$^{1,}$\footnote{wangqi@sxie.edu.cn; 18303476022@163.com}, Fu-Hu Liu$^{2,}$\footnote{Correspondence:fuhuliu@163.com; fuhuliu@sxu.edu.cn}, Khusniddin K. Olimov$^{3,4}$\footnote{Correspondence:khkolimov@gmail.com; kh.olimov@uzsci.net}}

\affiliation{$^1$College of Basic Courses, Shanxi Institute of 
Energy, Jinzhong 030600, China
\\
$^2$Institute of Theoretical Physics, State Key Laboratory
of Quantum Optics and Quantum Optics Devices \& Collaborative
Innovation Center of Extreme Optics, Shanxi University,
Taiyuan 030006, China
\\
$^3$Physical-Technical Institute of Uzbekistan Academy of
Sciences, Chingiz Aytmatov str. 2b, Tashkent 100084, Uzbekistan
\\
$^4$National University of Science and Technology MISIS
(NUST MISIS), Almalyk Branch, Almalyk, Uzbekistan}

\begin{abstract}

\vspace{0.5cm}

\noindent {\bf Abstract:} The squared momentum transfer spectra of $\eta$ and
$\eta ^0$, produced in high-energy photon-proton ($\gamma p$)
$\rightarrow \eta(\eta^0)+p$ processes in electron-proton ($ep$)
collisions performed at CEBAF, NINA, CEA, SLAC, DESY, and WLS are
analyzed. The Monte Carlo calculations are used in the analysis of
the squared momentum transfer spectra, where the transfer
undergoes from the incident $\gamma$ to emitted $\eta(\eta^0)$ or
equivalently from the target proton to emitted proton. In the
calculations, the Erlang distribution and Tsallis-Levy function
are used to describe the transverse momentum ($p_T$) spectra of
emitted particles. Our results show that the average transverse
momentum ($\langle p_T\rangle$), the initial-state temperature
($T_i$), and the final-state temperature ($T_0$) roughly decrease
from the lower center-of-mass energy ($W$) to the higher one in
the concerned energy range of a few GeV, which is different from
the excitation function from heavy-ion collisions in the similar
energy range.

\end{abstract}
\pacs{12.40.Ee, 14.40.-n, 24.10.Pa, 25.75.Ag \vspace{0.5cm}}

\maketitle

\section{Introduction}

Abundant experimental data produced at the Large Hadron Collider
(LHC) and Relativistic Heavy Ion Collider (RHIC) are helpful for
scientists to study production of Quark-Gluon Plasma (QGP) and the
evolution of the collision system. In the process of high-energy
heavy-ion collisions, the time evolution of the collision system
roughly consists of five stages which are: flight of incoming
nuclei, beginning of collisions, strongly-coupled QGP (sQGP),
mixed phase, and hadron gas, respectively~\cite{2}. In each stage,
the evolution picture and property of the collision system, and
the distribution law and property of the produced particles, are
possibly different from others, because some particles are
produced in the earlier processes and others are produced in the
last stage.

In the initial stages, two nuclei with the shape of pancake due to
the Lorentz contraction move toward each other and collide
violently. Because of the transformation from the kinetic energy
of the particles to the huge amount of the thermal energy of the
system, after a short period of time ($\sim 1$ fm/$c$), QGP is
produced which is the extremely hot and dense
matter~\cite{3,4,5,6}. In the stages of mixed phase and hadron
gas, due to the inflation and cooling down of the system, the
hadron matter evolves until only color-neutral states are created.
In high-energy collisions, the excitation and equilibrium degrees
of the system are considered as the important characteristics
which can help us to study the mechanism of the nuclear reaction
and the characteristics of the system
evolution~\cite{7,8,9,10,11,12,13,14,15,16}.

In the whole process of high-energy collisions, one can use
different temperatures to describe the excitation degree of the
system or emission source at different
stages~\cite{17,18,19,20,21,22,23,24}. At the first place, one can
choose the initial-state temperature ($T_i$) to describe the
excitation degree of the system at the beginning of collisions. At
the second place, one can use the critical temperature ($T_c$) and
the chemical freeze-out temperature ($T_{ch}$) to describe the
excitation degree of the system in which the hadron matter appears
and chemical freeze-out happens separately. At the last place, one
can use the kinetic freeze-out or final-state temperature
($T_{kin}$ or $T_0$) and the effective temperature ($T_{eff}$) to
describe the excitation degree of the system at the kinetic
freeze-out. Here, $T_{eff}$ includes in addition the flow effect
and can be compared with $T_{kin}$ or $T_0$.

As a useful tool for describing the excitation degree of the
system, $T_i$ represents the temperature of the system or emission
source at the initial-stage of collisions~\cite{24a,24b}. This
initial-stage refers to a very short stage after the
thermalization at the beginning of collisions. To obtain $T_i$, we
have several methods. The first method is to solve the state
equation of QGP with fluid model~\cite{20}. The second one is to
solve the equation for isentropic expansion in relativistic fluid
mechanics~\cite{17}. The last one is to use the transverse
momentum ($p_T$) spectra directly, or use various distributions or
functions to fit the $p_T$ spectra. The last method has special
advantages of accuracy and efficiency. This is because there is no
need to study the concrete evolution process from QGP or sQGP to
hadron phase, but analyze the $p_T$ spectra themselves. Usually,
we can use the Erlang distribution~\cite{25,26,27}, Hagedorn
function~\cite{28}, and Tsallis-Levy function~\cite{29} for
fitting the $p_T$ spectra to obtain $T_i$, but in this paper, only
the Erlang distribution is selected due to it being the origin of
multiple sources in the multi-source thermal
model~\cite{25,26,27}.

The final-state temperature $T_0$ known as the kinetic freeze-out
temperature represents the temperature of the system or emission
source at the kinetic freeze-out stage. At this stage, the
interactions between various particles are negligible, and there
is no further elastic collisions in the system. $T_0$ can be
extracted by using the certain distribution or function in fitting
the $p_T$ spectra, and used to describe the excitation degree of
the system. $T_{eff}$ is similar to $T_0$, but there is no
influence of flow effect in $T_0$. In our previous
work~\cite{24a,24b}, we have used the Tsallis-Levy
function~\cite{29} in fitting the $p_T$ spectra to estimate $T$ as
$T_{eff}$. In some small systems such as $\gamma p$ and $\gamma^*
p$ collisions, we have $T_0 \approx T$, because the flow effect
are negligible. In the $\gamma p$ collisions discussed in this
work, we use the Tsallis-Levy function~\cite{29} in fitting the
$p_T$ spectra to extract $T$ ($T_{eff}$) as $T_0$ and study the
characteristics of the system at the last stage. As for the other
temperature types, we do not discuss them in our work anymore.

One can use some parameters to describe the equilibrium degree of
the system. By fitting the $p_T$ spectra with the Tsallis
distribution~\cite{31,32}, the entropy index $q$ is extracted.
Generally, if $q$ is closer to 1, the system is closer to
equilibrium, or the equilibrium degree of the system is higher. In
the absence of the Tsallis distribution~\cite{31,32}, one can use
the Hagedorn function~\cite{28} or Tsallis-Levy function~\cite{29}
alternatively. In the fitting process, $q$ can be abstracted by
introducing $n$, which is used to describe the equilibrium degree
of the system indirectly due to $n=1/(q-1)$. In this work, let $n$
be fixed, then we may obtain $T_i$ and $T_0$ more conveniently to
describe the excitation degree of the system. It should be noted
that when considering the equilibrium issue, small systems are
also possible because we consider a large number of events within
the framework of giant canonical ensemble.

To obtain the above-mentioned temperatures and $\langle
p_T\rangle$, we need to use the $p_T$ spectra. In the absence of
the $p_T$ spectra, we can use the squared momentum transfer
($|t|$) spectra alternatively. The squared momentum transfer is
one of the Mandelstam variables which consists of the
four-momentum of the concerned particles~\cite{33}. In the fitting
process, the squared momentum transfer spectra can not be fitted
by those distributions and functions directly. Instead, we can
obtain many concrete $p_T$ satisfying certain distribution, then
we can obtain many concrete values of the squared momentum
transfer with the Monte Carlo calculation. Finally, we can obtain
the distribution of the squared momentum transfer.

In this paper, the squared momentum transfer spectra of $\eta$ and
$\eta^0$, produced in high-energy $\gamma p$ collisions performed
at the Continuous Electron Beam Accelerator Facility
(CEBAF)~\cite{34}, the Daresbury Laboratory electron synchrotron
NINA~\cite{35}, the Cambridge Electron Accelerator
(CEA)~\cite{36}, the Stanford Linear Accelerator Center
(SLAC)~\cite{37}, the Deutsches Elektronen-Synchrotron
(DESY)~\cite{38}, and the Wilson Laboratory Synchrotron
(WLS)~\cite{39} are fitted by the results obtained with the Monte
Carlo method. These experimental data are measured at different
center-of-mass energies ($W$) and incident photon energies
($E_\gamma$).

The remainder of this article is structured as follows. The
formalism and method are described in Section 2. Results and
discussion are given in Section 3. In section 4, we give our
summary and conclusions.

\section{Picture and formalism}

i) {\it The Erlang distribution}

The Erlang distribution which describes the $p_T$ spectra and
multiplicity distribution can be obtained from the multi-source
thermal model~\cite{25,26,27}, where the multiplicity is defined
as the number of particles produced in an event. The model assumes
that multiple sources are formed and contribute to $p_T$ of
considered particles in collision process. These sources are
considered as nucleons or partons if we study the formation of
nucleon clusters (nuclear fragments) or particles. Generally
speaking, it is enough to use one or two-component Erlang
distribution in fitting the $p_T$ spectra.

The Erlang distribution is the convolution of multiple exponential
distributions~\cite{25,26,27}. Every exponential distribution
represents the transverse momentum ($p_t$) distribution obeyed by
a parton, and can be regarded as
\begin{align}
f(p_{tj})=\frac{1}{\langle p_t \rangle}
\exp\bigg(-\frac{p_{tj}}{\langle p_t \rangle}\bigg).
\end{align}
Here, $j$, $p_{tj}$, and $\langle p_t\rangle$ refer to the index
of participant partons, the transverse momentum which depends on
$j$, and the average contribution of participant partons to
$\langle p_T \rangle$ of the considered particles, respectively.

It is assumed that $n_s$ partons contribute to $p_T$ of a given
particle. We have the Erlang $p_T$ distribution to be
\begin{align}
f_1(p_T)=\frac{1}{N}\frac{dN}{dp_T}=\frac{p_T^{n_s-1}}{(n_s-1)!{\langle
p_t \rangle}^{n_s}} \exp\bigg(-\frac{p_T}{{\langle p_t
\rangle}}\bigg).
\end{align}
In Eq. (2), the $p_T$ of a given particle consists of $p_{t1}$,
$p_{t2}$, ..., $p_{tn_s}$ of $n_s$ partons. Here $n_s$ is not
large and it is around 2--5. This is because $n_s$ is not
determined by the collision system, but by the number of partons
contributing to a given $p_T$. As for $N$, it is the number of
particles, and it depends on the collision system. It is natural
that $\int_0^{\infty} f_1(p_T) dp_T=1$ because $f_1(p_T)$ is a
probability density function.
\\

ii) {\it The Tsallis-Levy function}

The Tsallis-Levy function is one of the applications of the
Tsallis statistics~\cite{31} in high-energy collisions. We have
$p_T$ distribution in form of the Tsallis-Levy function~\cite{29}
to be
\begin{align}
f_2(p_T)=\frac{1}{N}\frac{dN}{dp_T}
=Cp_T\bigg(1+\frac{\sqrt{p_T^2+m_0^2}-m_0}{nT}\bigg)^{-(n+1)}.
\end{align}
Here $T$ and $n$ are free parameters, $\sqrt{p_T^2+m_0^2}=m_T$ is
the transverse mass, $m_0$ is the rest mass of the considered
particle, and $C$ is the normalization constant which is related
to $T$, $n$, and $m_0$ to make $\int_0^{\infty} f_2(p_T) dp_T=1$.
Due to particle mass $m_0$ appearing in $\sqrt{p_T^2+m_0^2}-m_0$
in Eq. (3), $f_2(p_T)$ is related to $m_0$. Our tentative
calculation shows that $m_0$ affects mainly the normalization and
weakly the tendency of the function.

In the fitting process of the $p_T$ spectra with the Tsallis-Levy
function~\cite{29}, we can obtain $T$ which is used to describe
the excitation degree of system at the kinetic freeze-out stage.
The influence of flow effect is included in $T$ compared with
$T_0$. In general, $T \textgreater T_0$, but in the $\gamma p$
collision discussed in this work, due to the flow effect being
small and considered negligible, we are of the opinion that $T
\approx T_0$ roughly. To obtain the excitation function of $T$
more conveniently, we set $n$ as a fixed value in the fitting with
the Tsallis-Levy function~\cite{29}.
\\

iii) {\it Average transverse momentum and initial-state
temperature}

In the process of fitting $p_T$ spectra with the Erlang
distribution~\cite{25,26,27}, $\langle p_T\rangle$ and $T_i$ are
estimated and used to describe the excitation degree of the
system. In fact, $\langle p_T\rangle$ can be obtained by
\begin{align}
\langle p_T\rangle =\int_0^{\infty} p_T f_1(p_T) dp_T =n_s\langle
p_t\rangle.
\end{align}
To obtain $T_i$, we need to use a color string percolation
method~\cite{40,41,42} which gives
\begin{align}
T_i =\sqrt{\frac{\langle p_T^2 \rangle}{2F(\xi)}},
\end{align}
where
\begin{align}
\langle p_T^2\rangle =\int_0^{\infty} p^2_T f_1(p_T)
dp_T=n_s(n_s+1)\langle p_t\rangle^2
\end{align}
and $F(\xi)$ is the color suppression factor. Although $f_2(p_T)$
can be also used in the calculation of $\langle p_T\rangle$ and
$\langle p_T^2\rangle$, it is more convenient to use $f_1(p_T)$
from which specific results for $\langle p_T\rangle$ and $\langle
p_T^2\rangle$ can be obtained from the integration.

It is necessary to discuss the application of color string
percolation method. In the process of using this method, we can
determine the number of strings used. For instance, only one
string is used in present work, that is to say
$F(\xi)=1$~\cite{43}. If we consider other strings, there will be
the minimum $F(\xi)\approx0.6$ which results in the maximum
increase of 29.1\% in $T_i$~\cite{43}. Although it is possible to
have any other strings in this work, they do not have a great
influence on $T_i$. This is because one string accounts for a
large proportion, but two and multiple strings account for a small
one.
\\

iv) {\it The squared momentum transfer}

In the center-of-mass reference frame, in two-body reaction
$2+1\rightarrow 4+3$ or two-body-like reaction, three Mandelstam
variables~\cite{33}, $s$, $t$, and $u$ are defined. They are
composed of four-momentum of participated particles and their
forms are
\begin{align}
s=-({P_1}+{P_2})^{2}=-({P_3}+{P_4})^{2},
\end{align}
\begin{align}
t=-({P_1}-{P_3})^{2}=-(-{P_2}+{P_4})^{2},
\end{align}
and
\begin{align}
u=-({P_1}-{P_4})^{2}=-(-{P_2}+{P_3})^{2},
\end{align}
respectively. Here, $P_{1}$, $P_{2}$, $P_{3}$, and $P_{4}$ are
four-momenta of particles 1, 2, 3, and 4, separately. Particle 1
is the target proton which is supposed to be incident along the
$Oz$ direction, and particle 2 is the incident $\gamma$ which is
supposed to be incident along the opposite direction. After
collisions, particle 3 is the emitted proton which is emitted with
angle $\theta$ relative to the $Oz$ direction, and particle 4 is
the emitted meson which is emitted along the opposite direction.

Due to different forms of Mandelstam variables, the physical
meanings of $s$, $t$, and $u$ are different. $\sqrt{s}$ is
supposed to be the center-of-mass energy, both $-u$ and $-t$ refer
to the squared momentum transfer between particles. In this work,
we choose variable $-t$ to research, and its form is
\begin{align}
|t|=|({E_1}-{E_3})^{2}-({\vec{p}_{1}}-{\vec{p}_{3}})^{2}|
=\bigg|m_1^2+m_3^2-2{E_1}\sqrt{\bigg({\frac{p_{3T}}{\sin\theta}}\bigg)^{2}+m_3^2}
& +2\sqrt{E_1^2-m_1^2}\frac{p_{3T}}{\tan\theta}\bigg|,
\end{align}
where $E_1$ and $E_3$, $\vec{p}_1$ and $\vec{p}_3$, as well as
$m_1$ and $m_3$ are the energy, momentum, and rest mass of
particles 1 and 3 respectively. Besides, $p_{3T}$ is the
transverse momentum of particle 3 which obeys Eq. (2) or (3).

In this paper, we select the squared momentum transfer spectra at
different center-of-mass energy $W$ and incident photon energy
$E_\gamma$ to analyze. The center-of-mass energy is $W= \sqrt{s}
=\sqrt{-(P_1 +P_2)^2}$ in our analysis~\cite{24a,24b}. Let $Q^2$
and $x_B$ be the squared photon virtuality and Bjorken scaling
variable, and we have $W^2 \simeq Q^2
/x_B$~\cite{44,45,46,47,48,49,50,51}.
\\

v) {\it The process of Monte Carlo calculations}

Although we can use Eq. (10) to obtain the single squared momentum
transfer, its distribution is difficult to obtain. To obtain the
squared momentum transfer distribution, we can execute the
following steps. At first, we produce many concrete $p_{3T}$
satisfied with Eq. (2) or (3) and $\theta$. At second, we can
obtain many concrete squared momentum transfer by calculating with
Eq. (10) repeatedly. At last, the squared momentum transfer
distribution is obtained with the statistical method.

To produce many concrete $p_{3T}$ and $\theta$, we may use the
Monte Carlo method. Let $R_{1,2}$ and $r_{1,2,3,...,n_s}$ be
random numbers distributed evenly in $[0,1]$. Then, we obtain many
concrete $p_{3T}$ by solving this equation
\begin{align}
\int_0^{p_T}f\left(p'_T\right)dp'_T<R_1<\int_0^{p_T+\delta
p_T}f\left(p'_T\right)dp'_T,
\end{align}
where $\delta p_T$ is a small shift relative to $p_T$, and
$f(p'_T)$ represents Eq. (2) or (3). As for Eq. (2), there is a
simpler expression of $p_T$. We can solve the equation
\begin{align}
\int_0^{p_{tj}}f\left(p'_{tj}\right)dp'_{tj}=r_j\qquad(j=1,2,3,...,n_s),
\end{align}
which results in
\begin{align}
p_{tj}=-\langle p_t\rangle \ln r_j\qquad(j=1,2,3,...,n_s).
\end{align}
In this way, the simpler expression is written as
\begin{align}
p_{T}=\sum_{j=1}^{n_s} p_{tj} =-\langle p_t\rangle
\sum_{j=1}^{n_s} \ln r_j=-\langle p_t\rangle
\ln\bigg(\prod_{j=1}^{n_s} r_j\bigg).
\end{align}

The distribution of $\theta$ satisfies with
\begin{align}
f_\theta\left(\theta\right)= \frac{1}{2}\sin \theta
\end{align}
which is the half-sine function. In the source's rest frame, it is
obtained under the assumption of isotropic emission. Solving the
equation
\begin{align}
\int_0^{\theta}f_\theta\left(\theta'\right)d\theta'=R_2,
\end{align}
we have
\begin{align}
\theta=2\arcsin \left(\sqrt{R_2}\right)
\end{align}
which is used in our calculations.

The squared momentum transfer distribution obtained using the
above steps is used to fit the experimental data measured at
different $W$ and $E_\gamma$. In the fitting process, parameters
$\langle p_t\rangle$, $n_s$, $T$ are extracted with the method of
least squares, and $n$ is fixed to be large enough for
convenience. Then, we can obtain $\langle p_T\rangle$ from Eq. (4)
and $T_i$ from Eq. (5). The errors of parameters are obtained by
the general method of statistical simulation.

\section{Results and discussion}

Figure 1 shows the differential cross-section, $d\sigma/d|t|$, in
the squared momentum transfer $|t|$ of $\gamma p \rightarrow \eta
p$ produced in different center-of-mass energy ranges
$2.52<W<2.56$, $2.60<W<2.64$, $2.64<W<2.68$, $2.68<W<2.72$,
$2.72<W<2.76$, $2.76<W<2.80$, $2.80<W<2.84$, $2.84<W<2.88$,
$2.88<W<2.92$, $2.92<W<2.96$, $2.96<W<3.00$, $3.04<W<3.08$, and
$3.08<W<3.12$ GeV, corresponding to the incident photon energy
range $E_\gamma \in \left[ 2.91, 4.72\right]$ GeV. The black
squares represent the experimental data performed at the CEBAF and
measured by the CEBAF Large Acceptance Spectrometer (CLAS)
Collaboration~\cite{34}, where the data in $2.56<W<2.60$ and
$3.00<W<3.04$ GeV are not available from the experiment. The green
solid curves and red dash-dotted curves are the statistical
results of $|t|$ in which $p_T$ satisfies the Erlang distribution
and Tsallis-Levy function, respectively. One can see that the
fitting results are in agreement with the experimental data.

\begin{figure*}[htbp]
\begin{center}
\rotatebox[origin=c]{-90}{\includegraphics[width=22cm]{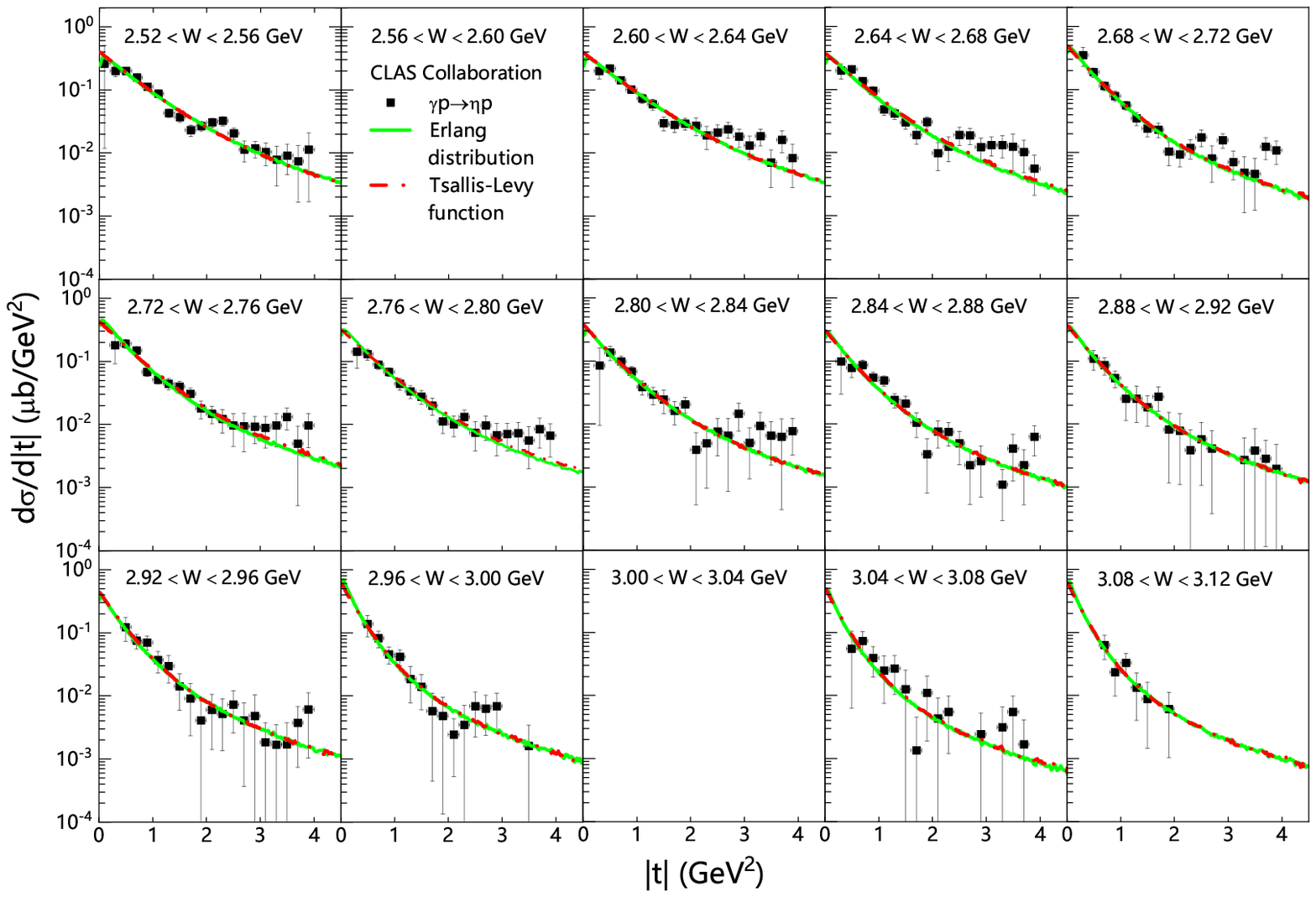}}
\end{center} \vspace{5mm}
\justifying\noindent {\small Figure 1. The differential cross-section $d\sigma/d|t|$
in $|t|$ of $\gamma p \rightarrow\eta p$ process produced in $ep$
collisions at energy ranges shown in the panels. The symbols
represent the experimental data measured by the CLAS
Collaboration~\cite{34}, where the data in $2.56<W<2.60$ and
$3.00<W<3.04$ GeV are not available from the experiment. The green
solid curves and red dash-dotted curves are the statistical
results of $|t|$ in which $p_T$ satisfies the Erlang distribution
and Tsallis-Levy function, respectively.}
\end{figure*}

\begin{figure*}[htbp]
\begin{center}
{\includegraphics[width=15cm]{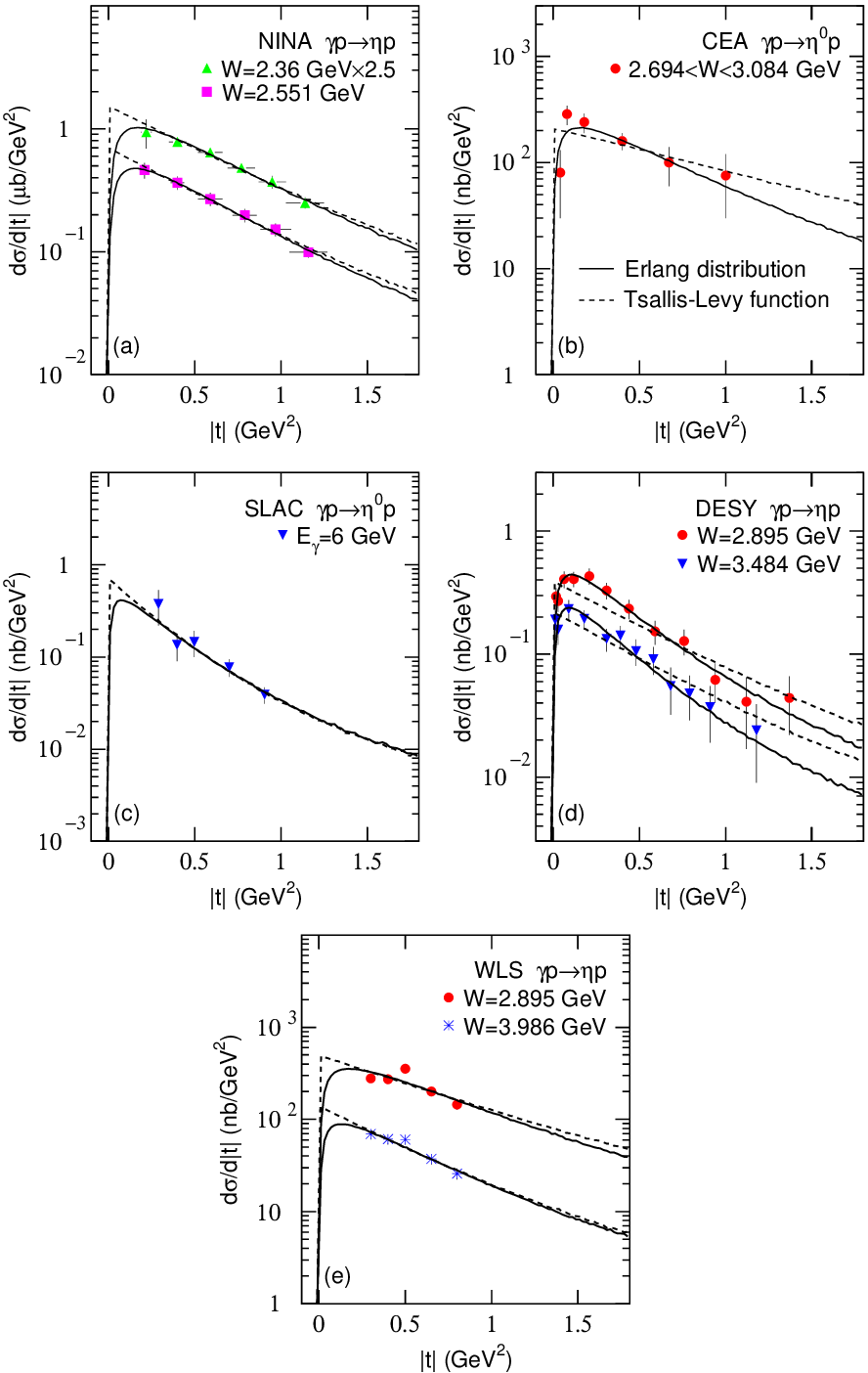}}
\end{center}
\justifying\noindent {\small Figure 2. The differential cross-section $d\sigma/d|t|$
in $|t|$ of (a, d, e) $\gamma p \rightarrow \eta p$ and (b, c)
$\gamma p \rightarrow \eta^0 p$ process produced at (a)
NINA~\cite{35}, (b) CEA~\cite{36}, (c) SLAC~\cite{37}, (d)
DESY~\cite{38}, and (e) WLS~\cite{39} at different $W$ and
$E_\gamma$ shown in the panels. The symbols represent the
experimental data~\cite{35,36,37,38,39}. The black solid curves
and black dashed curves are the statistical results of $|t|$ in
which $p_T$ satisfies the Erlang distribution and Tsallis-Levy
function, respectively.}
\end{figure*}

\begin{table*}[!htb]
\vspace{.0cm} \justifying\noindent {\small Table 1. Values of $\langle p_t\rangle$, $n_s$, $T_i$, $T$, and
the first and last $\chi^2$/ndof corresponding to the statistical
results of $|t|$ in which $p_T$ satisfies the Erlang distribution
and Tsallis-Levy function, respectively, where $E_\gamma$ is used
for Figure 2(c) and $W$ is used for other cases in Figures 1 and
2.} \vspace{-4mm}
\begin{center}
\setlength\tabcolsep{6pt}%
{\small
\begin{tabular} {cccccccc}\\ \hline Figure & $W$, $E_\gamma$ (GeV) & $\langle p_t\rangle$ (GeV/$c$) & $n_s$ & $T_i$ (GeV) & $\chi^2$/ndof & $T$ (GeV)& $\chi^2$/ndof\\
\hline
Figure 1 & $(2.52,2.56)$ & $0.231\pm0.010$ & $3$ & $0.566\pm0.024$ & $32.76/17$ & $0.204\pm0.015$ & $33.62/18$\\
         & $(2.60,2.64)$ & $0.233\pm0.013$ & $3$ & $0.571\pm0.032$ & $22.93/16$ & $0.208\pm0.015$ & $23.63/17$\\
         & $(2.64,2.68)$ & $0.212\pm0.014$ & $3$ & $0.519\pm0.035$ & $27.79/16$ & $0.189\pm0.014$ & $26.76/17$\\
         & $(2.68,2.72)$ & $0.185\pm0.009$ & $3$ & $0.453\pm0.022$ & $20.99/16$ & $0.154\pm0.011$ & $21.47/17$\\
         & $(2.72,2.76)$ & $0.192\pm0.008$ & $3$ & $0.470\pm0.019$ & $14.69/16$ & $0.174\pm0.012$ & $14.21/17$\\
         & $(2.76,2.80)$ & $0.197\pm0.011$ & $3$ & $0.483\pm0.026$ & $9.59/16$  & $0.178\pm0.009$ & $8.83/17$\\
         & $(2.80,2.84)$ & $0.193\pm0.010$ & $3$ & $0.473\pm0.024$ & $14.40/16$ & $0.156\pm0.015$ & $14.42/17$\\
         & $(2.84,2.88)$ & $0.180\pm0.010$ & $3$ & $0.441\pm0.025$ & $20.97/15$ & $0.145\pm0.014$ & $20.68/16$\\
         & $(2.88,2.92)$ & $0.178\pm0.008$ & $3$ & $0.436\pm0.020$ & $2.63/13$  & $0.143\pm0.009$ & $2.67/14$\\
         & $(2.92,2.96)$ & $0.169\pm0.009$ & $3$ & $0.414\pm0.022$ & $6.19/15$  & $0.126\pm0.010$ & $5.54/16$\\
         & $(2.96,3.00)$ & $0.139\pm0.007$ & $3$ & $0.340\pm0.017$ & $8.64/11$  & $0.104\pm0.010$ & $8.52/12$\\
         & $(3.04,3.08)$ & $0.138\pm0.010$ & $3$ & $0.338\pm0.024$ & $7.28/11$  & $0.097\pm0.013$ & $7.14/12$\\
         & $(3.08,3.12)$ & $0.133\pm0.006$ & $3$ & $0.326\pm0.015$ & $1.78/3$   & $0.092\pm0.010$ & $1.75/4$\\
\hline
Figure 2(a)& $2.360$         & $0.172\pm0.003$ & $4$ & $0.544\pm0.010$ & $0.96/3$  & $0.199\pm0.006$ & $1.78/4$\\
           & $2.551$         & $0.165\pm0.002$ & $4$ & $0.522\pm0.007$ & $0.70/3$  & $0.188\pm0.002$ & $1.39/4$\\
Figure 2(b)& $(2.694,3.084)$ & $0.165\pm0.008$ & $4$ & $0.522\pm0.026$ & $5.87/3$  & $0.300\pm0.040$ & $9.46/4$\\
Figure 2(c)& $6$ & $0.118\pm0.005$ & $4$ & $0.373\pm0.016$ & $1.47/2$ & $0.097\pm0.007$ & $1.47/3$\\
Figure 2(d)& $2.895$         & $0.137\pm0.005$ & $4$ & $0.433\pm0.016$ & $9.78/9$  & $0.190\pm0.020$ & $21.75/10$\\
           & $3.484$         & $0.129\pm0.006$ & $4$ & $0.408\pm0.019$ & $8.42/9$  & $0.182\pm0.013$ & $6.37/10$\\
Figure 2(e)& $2.895$         & $0.175\pm0.007$ & $4$ & $0.553\pm0.022$ & $12.21/2$ & $0.220\pm0.025$ & $13.92/3$\\
           & $3.986$         & $0.152\pm0.005$ & $4$ & $0.481\pm0.015$ & $2.61/2$  & $0.158\pm0.013$ & $2.99/3$\\
\hline
\end{tabular}}
\end{center}
\end{table*}

\begin{figure*}[htbp]
\begin{center}
{\includegraphics[width=15cm]{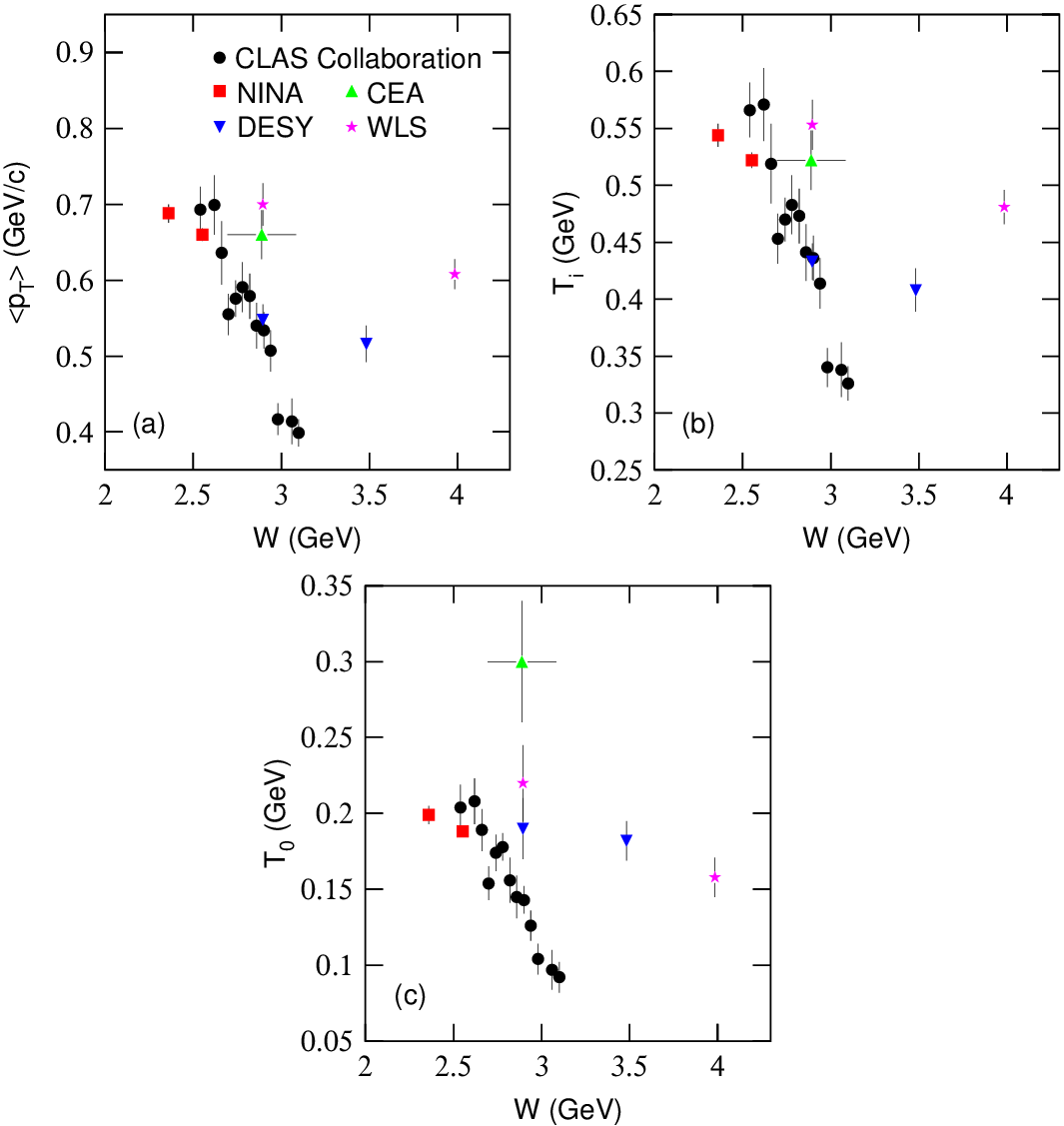}}
\end{center}
\justifying\noindent {\small Figure 3. The dependences of (a) $\langle p_T\rangle$,
(b) $T_i$, and (c) $T_0$ on $W$ in $ep$ collisions produced at
different devices.}
\end{figure*}

In the fitting process, the average transverse momentum $\langle
p_t\rangle$ contributed by participant partons, the number $n_s$
of participant partons, and the effective temperature $T$ of the
emission source are extracted. With the values of $\langle
p_t\rangle$ and $n_s$, the average transverse momentum $\langle
p_T\rangle$ of final-state particles and the initial-state
temperature $T_i$ of the emission source are obtained naturally.
To obtain $T_0$ more conveniently, we take $n=40$ ($q=1.025$) in
the Tsallis-Levy function~\cite{29}. Here, $T_0\approx T$ in the
function for the small system $\gamma p$ or $ep$ process in which
the effect of collective flow is neglected. In addition,
$E_1=0.938$ GeV in Eq. (10). In Table 1, we list the values of
free parameters $\langle p_t\rangle$ and $n_s$, derived parameter
$T_i$, and $\chi^2$/ndof for the fit of Erlang distribution, as
well as the values of free parameter $T$ and $\chi^2$/ndof for
the fit of Tsallis-Levy function.

Similar to Figure 1, Figure 2 presents the differential
cross-section, $d\sigma/d|t|$, in $|t|$ of (a, d, e) $\gamma p
\rightarrow \eta p$ and (b, c) $\gamma p \rightarrow \eta^0 p$
produced at (a) NINA~\cite{35}, (b) CEA~\cite{36}, (c)
SLAC~\cite{37}, (d) DESY~\cite{38}, and (e) WLS~\cite{39} at (a)
$W=$ 2.36, 2.551 GeV, in (b) 2.694 GeV $< W <$ 3.084 GeV, at (c)
$E_\gamma=$ 6 GeV, (d) $W=$ 2.895, 3.484 GeV, and (e) $W=$ 2.895,
3.986 GeV. The symbols in Figure 2 represent the experimental
data. The black solid and dashed curves are the statistical
results of $|t|$ in which $p_T$ satisfies the Erlang distribution
and Tsallis-Levy function, respectively. The values of parameters
and $\chi^2$/ndof are listed in Table 1. One can see that the
statistical results are in approximate agreement with the
experimental data.

The dependences of (a) $\langle p_T\rangle$, (b) $T_i$, and (c)
$T_0$ on center-of-mass energy ($W$) are given in Figure 3. The
different symbols represent the parameter values extracted from
Figures 1 and 2. For the results from the CLAS Collaboration, one
can see that $\langle p_T\rangle$, $T_i$, and $T_0$ decrease
generally with an increase in $W$. For the results from the other
cases, the trends are not clear. In the fitting process of
experimental data produced at CEA and DESY, we consider that the
ranges of $|t|$ are wider than others. To fit better, there is a
big difference between the statistical results of $|t|$ in which
$p_T$ satisfies the Erlang distribution and Tsallis-Levy function,
and it results in a higher $T_0$ extracted from the statistical
results of $|t|$ in which $p_T$ satisfies the Tsallis-Levy
function.

From Figure 3, it should be noted that the results for CEA and WLS
overlap for $\langle p_T\rangle$ and $T_i$, but differ
significantly for $T_0$, at $W\approx 3$ GeV. The reason is that
both $\langle p_T\rangle$ and $T_i$ are from the Erlang
distribution and $T_0$ is from the Tsallis-Levy function. In most
cases, the two fits are similar to each other. In a few cases
(Figures 2(b) and 2(d)), the two fits (the solid and dashed
curves) are inharmonious. If we try to obtain a similar result for
the two fits in a given range of $|t|$ (e.g. $|t|>0.2$ GeV$^2$), a
few data (the first or second one) will deviate greatly from the
fit. That is, we may adjust the parameters in Figure 3(c) to be
harmonious. However, a worse fit will be obtained. Due to the two
inharmonious fits in Figures 2(b), if the two fits of the Erlang
distribution in Figures 2(b) and 2(e) are harmonious (Figures 3(a)
and 3(b)), the two fits of the Tsallis-Levy function in the two
panels are inharmonious (Figure 3(c)). In addition, because two
parameters are used in the Erlang distribution and one parameter
is used in Tsallis-Levy function, the former is more flexible than
the latter in the fit.

Generally speaking, $\langle p_T\rangle$, $T_i$, and $T_0$
increase with the increase of $W$ in heavy-ion collisions at a few
GeV energy~\cite{11,18,19,20,21,22,24,43} which is the energy
range discussed in this work. Comparing with heavy-ion collisions,
in $\gamma p\rightarrow \eta p$ reaction the situation is
different due to the absence of secondary collision process and
cold nuclear effect in the small system. In addition, the small
system has not enough time to react at higher energy. This implies
that the small system has a lower excitation degree at higher
energy. Although this work confirms our previous work~\cite{24a},
the energy range discussed by us is narrow, and the data cited
here are measured at different devices with low statistics and
large errors in most cases. To obtain more solid an explicit
conclusions regarding the evolution of the parameters with
changing the energy, significantly higher statistics of the
experimental data are required in the future.

Before summary and conclusions, we would like to point out that
although the multiplicity is only two in the two-body reaction
discussed in the present work, we have used the parametrization
from the Erlang distribution and Tsallis-Levy function due to lots
of events being collected in experiments. This case can be
compared with the grand canonical ensemble in statistical physics.
Although the particles in different events do not have
interactions and the multiplicity in each event is very low, these
particles have the same or similar production condition due to the
same or similar events with given collision energy. Therefore, we
think that the particles in lots of events obey some statistical
laws. In addition, as a reflection of the average kinetic energy
of the thermal or disorganized motion, the concept of temperature
is applicable in the field of high energy collisions, even
two-body reaction.

\section{Summary and conclusions}

The squared momentum transfer spectra of $\eta$ and $\eta^0$
produced in the two-body process $\gamma p\rightarrow \eta(\eta^0)
+ p$ have been analyzed by the statistical results of $|t|$ in
which $p_T$ satisfies the Erlang distribution and Tsallis-Levy
function respectively. The squared momentum transfer undergoes
from the incident $\gamma$ to emitted $\eta$ or $\eta^0$, or also
equivalently from the target proton to emitted proton. The
statistical results are in agreement with the experimental data
measured at different experiments. In the fitting process, free
parameters $\langle p_t\rangle$, $n_s$, and $T$ are extracted.
Then, we obtain the dependences of $\langle p_T\rangle$, $T_i$,
and $T_0$ on center-of-mass energy $W$.

At a few GeV, it is believed that $\langle p_T\rangle$, $T_i$, and
$T_0$ increase generally with an increase in $W$ in heavy-ion
collisions. However, in $\gamma p\rightarrow \eta p$ reaction the
situation is different due to the absence of secondary collision
process and cold nuclear effect in the small system. Meanwhile,
the small system has not enough time to react at higher energy.
This implies that the small system has a lower excitation degree
at higher energy. The excitation functions of the concerned
parameters in the large and small systems have different
tendencies. More data are required in the future to compare the
excitation functions in the two kinds of systems.
\\
\\
\\
{\bf Author Contributions:} The authors contributed to the paper
in this way: conceptualization, F.-H.L. and K.K.O.; methodology,
F.-H.L. and K.K.O.; software, Q.W.; validation, F.-H.L. and
K.K.O.; formal analysis, Q.W.; investigation, Q.W.; resources,
Q.W.; data curation, Q.W.; writing -- original draft preparation,
Q.W.; writing -- review and editing, F.-H.L. and K.K.O.;
visualization, Q.W.; supervision, F.-H.L. and K.K.O.; project
administration, Q.W. and F.-H.L.; funding acquisition, Q.W.,
F.-H.L. and K.K.O. All authors have read and agreed to the
published version of the manuscript.
\\
\\
{\bf Funding:} The work of Q.W. was supported by the Shanxi
Provincial Natural Science Foundation under Grant No. 2023 and the
Doctoral Scientific Research Foundations of Shanxi Province and
Shanxi Institute of Energy. The work of F.-H.L. was supported by
the National Natural Science Foundation of China under Grant No.
12147215, the Shanxi Provincial Natural Science Foundation under
Grant No. 202103021224036, and the Fund for Shanxi ``1331 Project"
Key Subjects Construction. The work of K.K.O. was supported by the
Ministry of Innovative Development of the Republic of Uzbekistan
within the fundamental project No. F3-20200929146 on analysis of
open data on heavy-ion collisions at RHIC and LHC.
\\
\\
{\bf Institutional Review Board Statement:} Not applicable.
\\
\\
{\bf Informed Consent Statement:} Not applicable.
\\
\\
{\bf Data Availability Statement:} The data used to support the
findings of this study are included within the article and are
cited at relevant places within the text as references.
\\
\\
{\bf Conflicts of Interest:} The authors declare that there are no
conflicts of interest regarding the publication of this paper. The
funders had no role in the design of the study; in the collection,
analysis, or interpretation of the data; in the writing of the
manuscript; or in the decision to publish the results.
\\
\\
\end{document}